\documentclass[12pt,elsart,epsfig]{article}     % Paper without sections
\usepackage{epsfig}

\begin{document}
\begin{center}
{\Large \bf A partial wave analysis of $\bar pp \to \eta \eta \pi ^0$}
\vskip 5mm

{A.V. Anisovich$^c$, C.A. Baker$^b$, C.J. Batty$^b$, D.V. Bugg$^a$,
V.A. Nikonov$^c$, A.V. Sarantsev$^c$, V.V. Sarantsev$^c$, 
B.S.~Zou$^{a}$ \footnote{Now at IHEP, Beijinj 100039, China} \\
{\normalsize $^a$ \it Queen Mary and Westfield College, London E1\,4NS, UK}\\
{\normalsize $^b$ \it Rutherford Appleton Laboratory, Chilton, Didcot OX11 0QX,UK}\\
{\normalsize $^c$ \it St. Petersburg Nuclear Physics Institute, Gatchina, 
St. Petersburg district, 188350, Russia}\\ 
[3mm]}
\end {center}

\begin{abstract}
A partial wave analysis of $\bar pp \to \eta \eta \pi ^0$ data from the
Crystal Barrel experiment is made in terms of $s$-channel resonances.
The decay channels $a_0(980)\eta$, $f_0(1770)\pi$ and
$f_0(2105)\pi$ provide evidence for two $I = 1$ $J^{PC} = 0^{-+}$
resonances. The first has mass $M =2360 \pm 25$ MeV and width
$\Gamma = 300^{+100}_{-50}$ MeV, and the second $M =2070 \pm 35$ MeV,
$\Gamma = 310^{+100}_{-50}$ MeV.
There is also evidence for a $J^{PC} = 2^{-+}$ state with
$M = 2005 \pm 15$ MeV and $\Gamma = 200 \pm 40$ MeV, decaying strongly to
$a_0(980)\pi$.
\end{abstract}
\vskip 4mm

The present work is part of an analysis
of $\bar pp$ annihilation with isospin $I = 1$ and
charge conjugation number $C = +1$ in terms of $s$-channel resonances.
A combined analysis of data from final states $3\pi ^0$, $\pi ^0\eta$ and
$\pi ^0 \eta '$ is reported separately [1].
Here we focus attention on annihilation to  $\eta \eta \pi ^0$.
In these data, signals are visible from final states
$a_0(980)\eta$, $f_0(1500)\pi$, $f_0(1770)\pi$ and $f_0(2105)\pi$.
They carry no spin and may be analysed simply in
terms of Legendre polynomials describing the production process.

We find that the largest contributions to $\eta \eta \pi ^0$ data arise
from S-wave final states with $J^P = 0^-$;
those from $1^+$, $2^-$, $3^+$ and $4^-$ are somewhat smaller.
Hence these data give a rather direct determination of contributions
with quantum numbers $0^-$. In contrast, $3\pi ^0$ data contain
weak $0^-$ contributions from the final state $f_2(1270)\pi$
with orbital angular momentum $L = 2$ in the decay process.
There, the $0^-$ amplitude is hard to separate from larger
$L = 2$ $f_2(1270)\pi$ amplitudes with $J^P = 2^-$ and $4^-$.

%Fig. 1
\begin{figure}
\begin{center}
\vskip -35mm
\epsfig{file=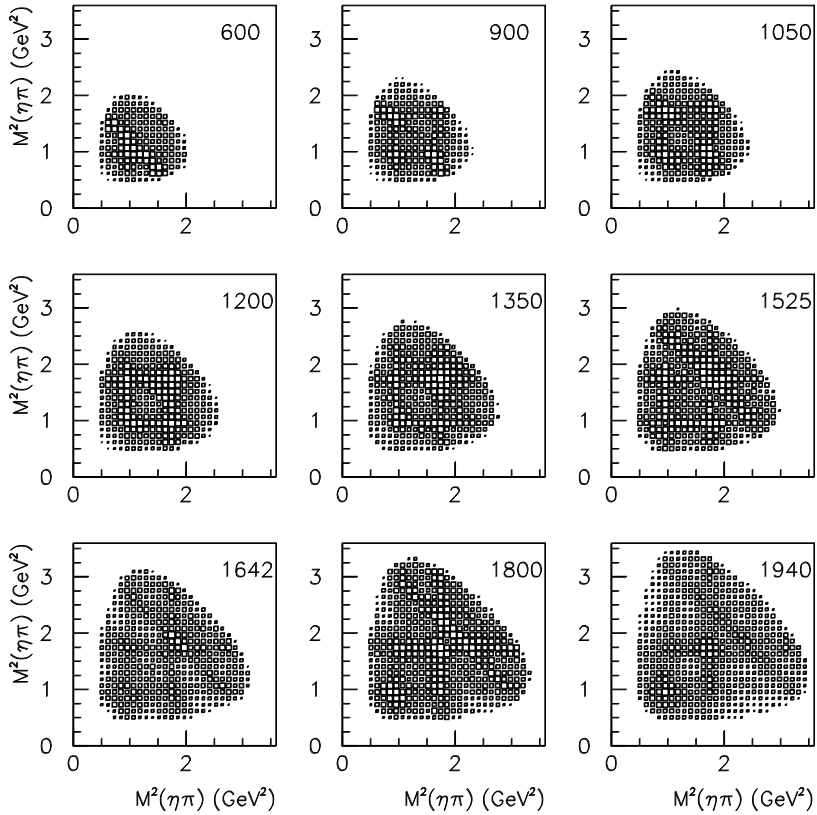,width=13.5cm}\
\vskip -135.45mm
\hskip 2.90mm
\epsfig{file=FEPE1.EPS,width=13.5cm}\
~\
\vskip -8mm
\caption{Dalitz plots for data; numbers in each panel indicate beam
momenta in MeV/c. }
\end{center}
\end{figure}

The $\eta \eta \pi ^0$ data treated here have been presented in two
earlier publications [2,3].
They may be fitted with the channels
\begin {eqnarray}
\bar pp &\to& a_0(980)\eta \\
        &\to& a_2(1320)\eta \\
        &\to& a_2(1660)\eta \\
        &\to& f_0(1500)\pi \\
        &\to& f_0(1770)\pi \\
        &\to& f_0(2105)\pi \\
        &\to& f_2(1980)\pi \\
        &\to& f_2(1270)\pi .
\end {eqnarray}
Fig. 1 shows Dalitz plots for data at all nine beam momenta. Figs. 2 and 3
show projections on to masses of $\pi \eta$ and $\eta \eta$
combinations; histograms show the fit described below.
There are clear peaks in Fig. 3 due to $f_0(1500) \to \eta \eta$,
and in Fig. 2 due to $a_0(980)$ and $a_2(1320) \to \pi \eta$.
The latter two are stronger than is immediately apparent from
the figure, since the $a_0$ and $a_2$ peaks originate from only one of
the two $\pi ^0\eta$ combinations, e.g. from $\pi ^0\eta _1$;
the other combination
$\pi ^0\eta _2$ produces a broad background when projected
on to $M(\pi ^0\eta _1)$, as one sees from Dalitz plots.

In Ref. [3], it was shown that a small but highly significant
peak in the $\eta \eta$ channel requires the presence of $f_0(1770)$
at beam momenta 900--1350 MeV/c.
In Ref. [2], it was also shown that data at 1525--1940 MeV/c require a
strong $f_0(2105) \to \eta \eta$ signal.
That resonance has also been observed in $\bar pp \to \eta \eta $[4]
and was first identified in $J/\Psi \to \gamma (4\pi )$ data [5],
where $f_0(2105) \to \sigma \sigma$, and $\sigma$ stands for the
$\pi \pi$ S-wave amplitude.
Ref. [2] also presented evidence for a broad $2^+$ signal in $\eta \eta$
with $M = 1980 \pm 50$ MeV, $\Gamma = 500 \pm 100$ MeV.
There is further evidence for this broad state in WA102 data on
central production of $4\pi$ [6].

%Fig. 2
\begin{figure}
\begin{center}
\vskip -27.5mm
\epsfig{file=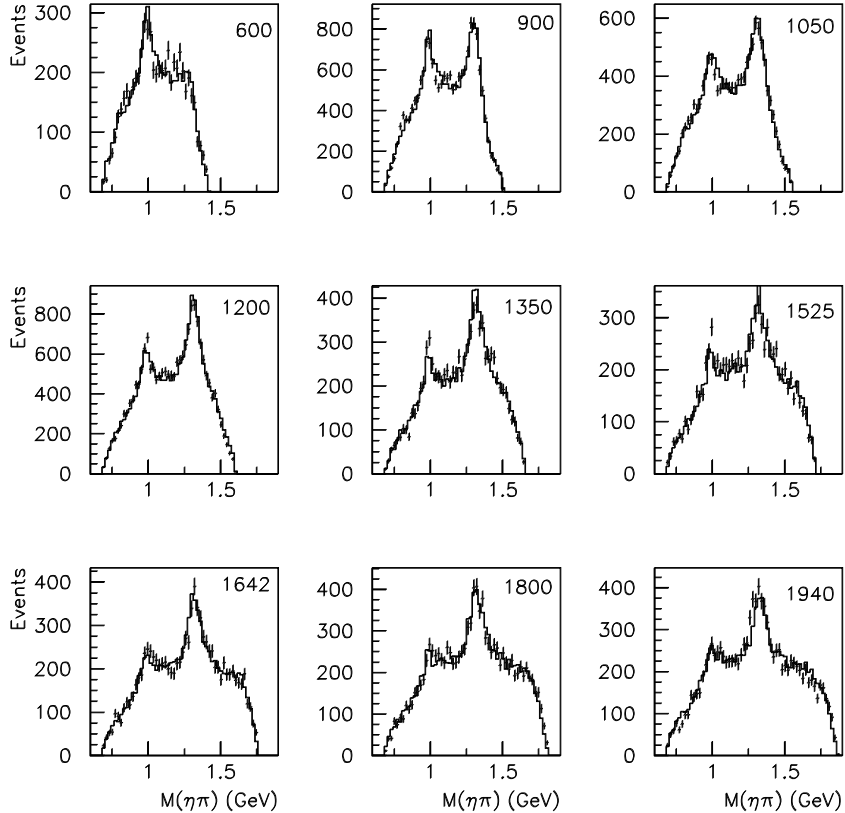,width=12.5cm}\
\vskip -125.45mm
\hskip 2.9mm
\epsfig{file=FEPE3.EPS,width=12.5cm}\
~\
\vskip -8mm
\caption{Projections on to $\eta \pi$ mass; histograms show
the partial wave fit. Numbers in each panel indicate beam momenta in
MeV/c.}
\end{center}
\end{figure}

We now describe the partial wave formula, given by equn. (9) below, 
used to fit the data.
This is the same as in the accompanying paper [1].
A full fit is made to production and decay of all channels (1)--(8)
in terms of $\bar pp$ partial waves up to $J^P = 4^-$.
Each partial wave has a distinctive dependence via relativistic
tensor expressions $Z(\theta, \alpha , \beta)$ on production angle
$\theta $ of each resonance in the centre of mass, and on its decay angles
$(\alpha, \beta ) $ in the rest frame of the resonance.
The energy dependence of each partial wave amplitude is expressed as the
sum of up to two $s$-channel resonances plus a background.
These resonances, described by simple Breit-Wigner functions
of $s$ with constant widths, are found to cluster in the mass
ranges 1930--2070 MeV and 2220--2360 MeV.
The background, where required, is taken as the
high energy tail of a resonance below the $\bar pp$ threshold.
This parametrisation guarantees that partial wave amplitudes
satisfy the necessary condition of analyticity, relating the
energy dependence of magnitudes and phases.

Blatt-Weisskopf centrifugal barrier factors $B_{\ell }(p)$ and
$B_L(q)$ are included in the partial wave amplitudes;
$B_{\ell }(p)$ incorporates the correct threshold dependence on
momentum $p$ in the $\bar pp$ channel for initial orbital angular momentum
$\ell$, and $B_L(q)$ likewise describes the dependence on $L$ and
momentum $q$ in the decay to the final state, eg. $a_0(980)\eta$.
The radius of the barrier is set to 0.83 fm from the determination
in Ref. [4].
The $a_0(980)$ amplitude for decay to particles 1 and 2 is
described by a Flatt\' e formula $F(s_{12})$ and other resonances
are described by Breit-Wigner amplitudes. In summary, the full
partial wave amplitude for channel (1), as an example, is given by
\begin {equation}
f = \frac {\sqrt {\rho _{\bar pp}}}{p} B_{\ell}(p)B_L(q)
\sum _i \frac {G_i}
{M_i^2 - s - iM_i\Gamma _i} [F(s_{12})Z_{12} + F(s_{23})Z_{23}],
\end {equation}
where $G_i$ are complex coupling constants and the sum $i$ is over
$s$-channel resonances and background.
Fitted parameters are $G_i$, $M_i$ and $\Gamma _i$.
The factor $1/p$ accounts for the flux in the entrance $\bar pp$ channel,
and $\rho _{\bar pp }$ is the phase space for this channel,
$2p/\sqrt {s}$.
Near the $\bar pp$ threshold, the S-wave cross section is then proportional
to $1/p\sqrt {s}$, the well-known $1/v$ law. In Fig. 4 below, fitted
cross sections will be shown multiplied by $ps^{1/2}$, so as to
display the resonant behaviour free of kinematic factors.
The phase space for the final state is accomodated in fitting the
Dalitz plot.

%Fig. 3
\begin{figure}
\begin{center}
\vskip -27.5mm
\epsfig{file=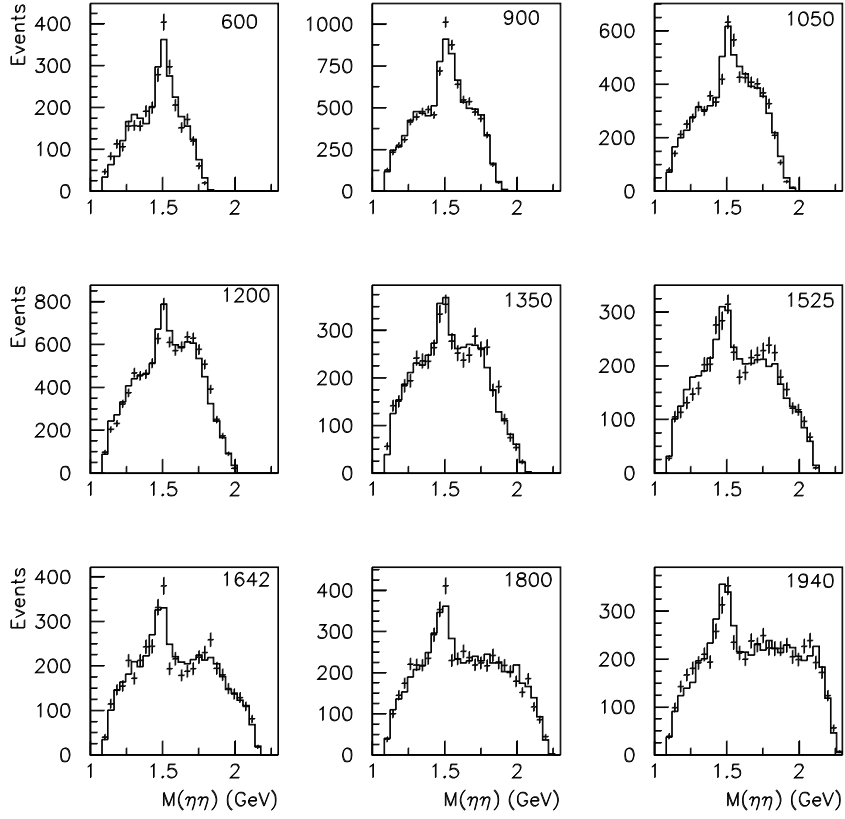,width=12.5cm}\
\vskip -125.45mm
\hskip 2.9mm
\epsfig{file=FEPE4.EPS,width=12.5cm}\
~\
\vskip -8mm
\caption{Projections on to $\eta \eta$ mass; histograms show the
partial wave fit. Numbers in each panel indicate beam momenta in MeV/c.}
\end{center}
\end{figure}

The parameters of each resonance are scanned in at least 7 steps
over a range of typically $\pm 100$ MeV.
For all quantum numbers other than $0^-$ and $2^-$, masses and
widths are determined less precisely by $\eta \eta \pi ^0$ data
than by $3\pi ^0$,
but are consistent with Ref. [1] within errors.
The $3\pi ^0$ data have statistics
$>100,000$ events per momentum whereas present data have
typically 5000--9000 events at each momentum.

%\vskip -20mm
\begin{table}[htp]
\begin{center}
\begin{tabular}{cccccc}
\hline
Name & $J^P$ & $M$   & $\Gamma$ & $M$ & $\Gamma $ \\
      &      & (MeV)    & (MeV) & (MeV) & (MeV)  \\\hline
$\pi $ & $0^-$  & $2070 \pm 35$  &  $310 ^{+100}_{-50}$ & $2070 \pm 35$ & $310
^{+100}_{-50}$ \\
$\pi $ & $0^-$  & $2355 \pm 25$  &  $270 ^{+100}_{-50}$ & $2360 \pm 25$ &
$300 ^{+100}_{-50}$ \\
$\pi _2$ & $2^-$  & $1990 \pm 30$  &  $290 \pm 60$ & $2005 \pm 15$ & $200 \pm
40$ \\
$\pi _2$ & $2^-$  &  -             &  -            & $2245 \pm 60$ &
$320 ^{+100}_{-40}$ \\
$\pi _4$ & $4^-$  & $2255 \pm 30$  &  $185 \pm 60$ & $2250 \pm 15$ &
$215 \pm 25$ \\\hline
$a_1 $ & $1^+$  & - & - & $1930 ^{+30}_{-70}$  &  $155 \pm 45$  \\
$a_1 $ & $1^+$  & - & - & $2270 ^{+55} _{-40}$  & $305 ^{+70}_{-40}$ \\
$a _2$ & $2^+$  & $2265 \pm 45$  &  $295 ^{+100}_{-60}$ & $2255 \pm 20$ &
$230 \pm 15$  \\
$a_3$ & $3^+$  &        -        &        -       & $2031 \pm 12$  &
$150 \pm 18$   \\
$a_3$ & $3^+$  & $2260 \pm 50$  &  $250 ^{+100}_{-50}$ & $2275 \pm 35$
& $350 ^{+100}_{-50}$  \\\hline
\end {tabular}
\caption {Columns 3 and 4 show masses and widths of resonances fitted to
$\eta \eta \pi ^0$ data.
Columns 5 and 6 show masses and widths from weighted averages
with fits to $3\pi ^0$, $\pi ^0 \eta$ and $\pi ^0 \eta '$.}
\end{center}
\end{table}

Columns 3 and 4 of Table 1 summarise masses and widths fitted to present data.
Errors cover systematic variations between decay channels; they also
cover variations depending on whether small components are included in the
fit or are omitted.
Table 2 gives a quantitative picture of the significance of each component
in the fit. It shows changes in log likelihood when each channel is
dropped and the remainder are re-optimised.
Our definition of log likelihood is such that it changes by 0.5 for
a one standard deviation change in one parameter.
Hence a change in log likelihood of 20 is rather significant
($\sim 5\sigma$, bearing in mind the number of fitted parameters).

%Fig. 4
\begin{figure}
\begin{center}
\vskip -30mm
\epsfig{file=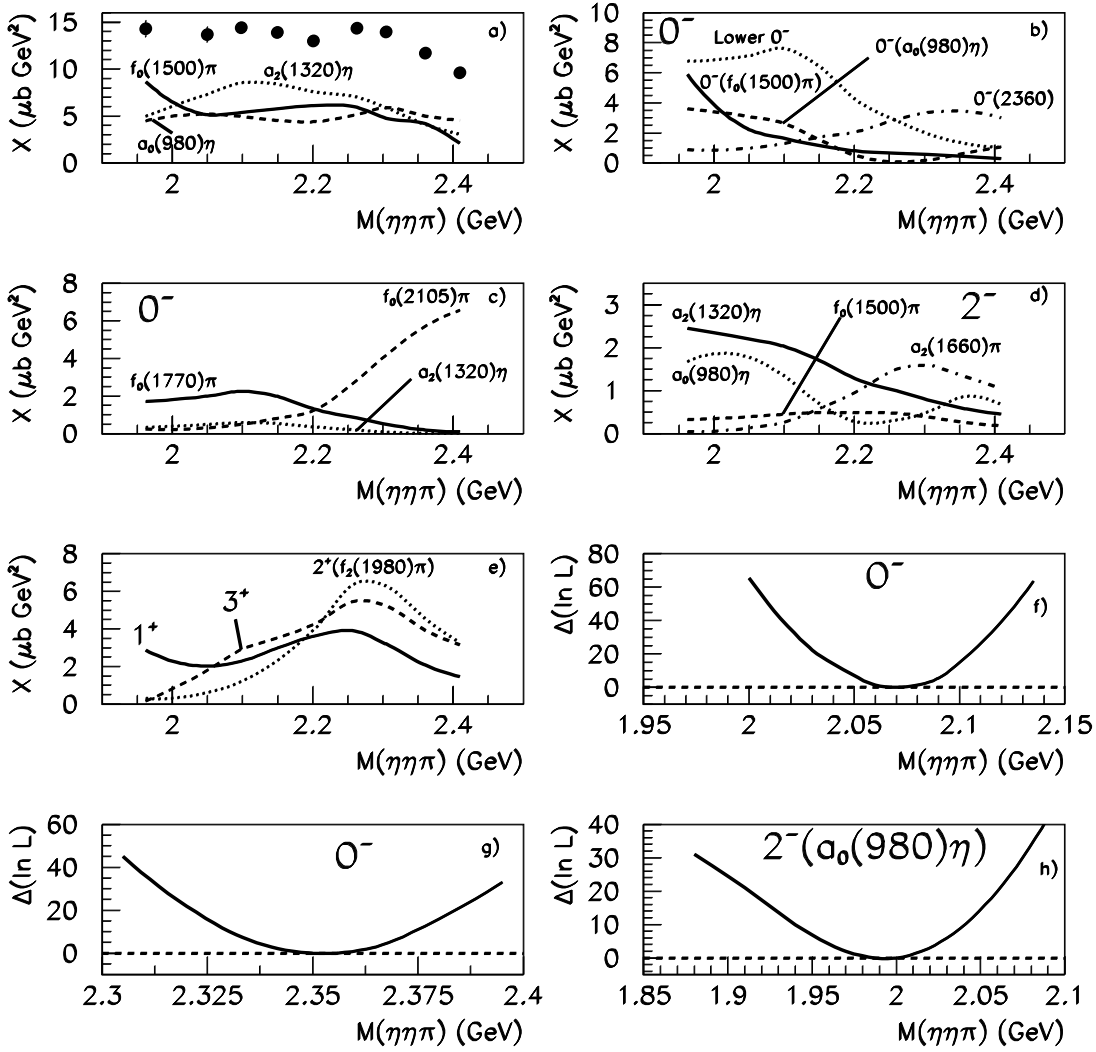,width=16.0cm}\
\vskip -160.45mm
\hskip 2.9mm
\epsfig{file=FEPE5.EPS,width=16.0cm}\
~\
\vskip -8mm
\caption {Values of $X = $ cross sections multiplied by $ps^{1/2}$ for
individual  channels.
(a) black circles show the overall cross section scaled by 1/3 for
comparison with summed contributions to $f_0(1500)\pi$,
$a_2(1320)\eta$ and $a_0(980)\eta$ from all $J^P$.
(b) $0^-$ contributions: the full curve is the summed
intensity from $f_0(1500)\pi$; remaining curves refer to
$a_0(980)\eta$; the dashed curve is the combined
intensity from all resonances, the dotted curve the coherent sum of
$\pi (1800)$ and $\pi (2070)$; the chain curve refers to
$\pi (2360)$.
(c) and (d): $0^-$ and $2^-$ intensities from individual
channels.
(e): summed $1^+$ and $3^+$ signals and that from
$2^+ \to [f_2(1980)\pi ]_{L = 1}$.
(f), (g) and (h): variations of log likelihood $ln L$ with mass
for  $\pi (2070)$, $\pi (2360)$ and the low mass $\pi_2 \to a_0(980)\eta$
signal.}
\end {center}
\end {figure}

\begin{table}[htp]
\begin{center}
\begin{tabular}{cccccc}
\hline
Resonance & $a_0(980)\eta$ & $f_0(1500)\pi$ & $f_0(1770)\pi$ & $a_2(1320)\eta$
& $f_0(2105)\pi$ \\
\hline
$\pi (1800)$  & 11 & 82 & - & - \\
$\pi (2070)$  & 272 & 12 & 324 & 126 \\
$\pi (2360)$  & 197 & 44 & 9 & 33 & 434  \\
$\pi _2(1880)$  &  -  & 255 & - & 176 \\
$\pi _2(2005)$  & 531  & - & - & - \\
$\pi _2(2245)$  & 107  & 10 & - & 19 \\
$\pi _4(2250)$  & 213 & -   & - & -   \\\hline
$a_1(1930)$  &  -  & 22 & - & 15  \\
$a_1(2270)$  & 18  & 77 & - & 71  \\
$a_3(2031)$  & 15  & 164& -  & 185        \\
$a_3(2275)$  & 141 & 37 & -  &  23 \\
$a_2(1950)$  & -   & -  & -  & 41  \\
$a_2(2030)$  & -   & -  & -  & 48  \\
$a_2(2175)$  & -   & -  & -  & 3  \\\hline
\end {tabular}
\caption {Changes in log likelihood when each
resonance is dropped from the fit, and remaining contributions are
re-optimised. In addition $\pi _2(2250) \to a_2(1660)\pi$ produces a change
of 107 and $a_2(2255) \to f_2(1980)\pi$ a change of 363. }
\end{center}
\end{table}

The black circles on Fig. 4(a) show the integrated cross section divided
by a factor 3, for comparison with those of individual channels.
These individual contributions are lower
because strong constructive interference between channels contributes
positively to the integrated cross section.
There is a sizeable cross section for the final state $a_2(1320)\eta $,
see the dotted curve of Fig. 4(a). However, it does not yield new physics.
This channel comes largely from $\bar pp$
triplet states with $J^P = 1^+$, $2^+$ and $3^+$. Resonances
in these partial waves are better determinined by $3\pi ^0$ data,
where statistics are very high and there are strong decays to $f_2(1270)\pi$.
The contribution to present data from $J^P = 4^+$ is negligible, probably
because of the $L = 3$ centrifugal barriers for both production and decay.
There is a small (7\%) intensity from $f_2(1270)\pi ^0$, but it has
little effect on other fitted amplitudes.
We find no significant contribution from $f_2'(1525)$ or $f_2(1565)$ at
any momentum.

For $J^P = 0^-$, three resonances are included.
One, $\pi (1800)$, is below the $\bar pp$ threshold and may simulate physics
background.
Two further resonances at 2070 and 2360 MeV are needed and are sufficient
to provide a good fit to the data.
Figs. 4(b) and (c) show $0^-$ contributions to the cross section.

We discuss the highest $0^-$ state first.
It receives large contributions from $f_0(2105)\pi$, shown by the
dashed curve on Fig. 4(c),  and from $a_0(980)\eta$.
The $f_0(2105)$ appears only in S-wave production and is
very secure.
The $f_0(1500)\pi$ channel makes only a small contribution
with $J^P = 0^-$ from both resonances at 2360 and 2070 MeV,
shown by the full curve of Fig. 4(b).
Table 2 shows that $a_0(980)\eta$ makes decisive contributions to
both resonances at 2360 and 2070 MeV.
For the $a_0(980)\eta$ channel, there is large interference between the upper
and lower $0^-$ resonances; the combined contribution is shown by the
dashed curve of Fig. 4(b).
Despite this interference, contributions from
$a_0(980)\eta$ are very stable.
This channel and $f_0(2105)\pi$  agree closely on a mass $M = 2355 \pm 25$
MeV for present data.
The cross section for production of $f_0(2105)\pi$ rises at high mass faster
than phase space for that channel and requires production through
$\pi (2360)$.
Fig. 4(g) shows the variation of log likelihood as the mass is varied.
All fits with a variety of ingredients give masses in the range
2337 to 2377 MeV.
This range is used to assess the systematic error, which is much
larger than the statistical error.
The fit to $3\pi ^0$ data gives a higher but distinctly less accurate mass
$2385 \pm 45$ MeV. The weighted mean of $2360 \pm 25$ MeV fits both $\eta \eta
\pi ^0$ data and $3\pi ^0$ well.
The width is much less well determined: $\Gamma = 300 ^{+100}_{-50}$ MeV.

At lower masses, there is a sizeable contribution from
$a_0(980)\eta$  and also from $f_0(1770)\pi $, shown by the full curve
of Fig. 4(c).
In the previous analysis of Ref. [3], it was shown that $f_0(1770)$
production peaks in the momentum range 900 to 1200 MeV/c.
Table 2 shows that $f_0(1770)\pi$ makes a
highly significant contribution of 324 to log likelihood.
It is particularly useful, since
it is again produced only through the S-wave and does not
contribute to $\pi (1800)$, which is too low in mass.
The cross section for production of $f_0(1770)\pi$ does not follow
phase space for that channel, but requires production through the
resonance at 2070 MeV; there is very little contribution from the
upper $0^-$ state.

From the present $\eta \eta \pi ^0$ data, the optimum parameters of
the lower resonance are $M = 2070 \pm 35$ MeV, $\Gamma = 310 ^{+100}_{-50}$
MeV, where errors are mostly systematic.
Fig. 4(f) shows log likelihood against mass when the
background is fitted by $\pi (1800)$.
However, the possibility of alternative descriptions of the background
introduces a systematic error of $\pm 35$ MeV in the mass determination.
The $3\pi ^0$ data of Ref. [1] also require the presence of $\pi (2070)$,
but again suffer from interference with a background term.
From those data, the optimum is at $2090 \pm 65$ MeV; the
width is large and poorly determined, $285 \pm 75$ MeV.
Both sets of data are well fitted with a mass of $2070 \pm 35$ MeV.

The two $0^-$ states lie rather higher in mass than
corresponding $I = 0$ states [4]. These were observed at
$2010 ^{+35}_{-60}$ and $2285 \pm 20$ MeV. In Ref. [4], all states
were found to lie close to straight-line trajectories of
$M^2$ against radial excitation number, with an average slope of
$1.143 \pm 0.013$ GeV$^2$ per excitation.
The spacing of the two $0^-$ states observed here is consistent
with this empirical rule.
One cannot compare accurately with $\pi (1300)$
because of the large uncertainty in its mass.
The $\pi (1800)$ is consistent in mass with the required intermediate
state, but there is evidence from its decay modes in favour of
interpretation as a hybrid [7].
The VES group has also reported evidence in $\omega \rho$ for a $0^-$ state
at 1750 MeV; this is a favoured decay mode for $q\bar q$ states [7] and
may be the second radial excitation.

%Fig. 5
\begin{figure}
\begin{center}
\vskip -23mm
\epsfig{file=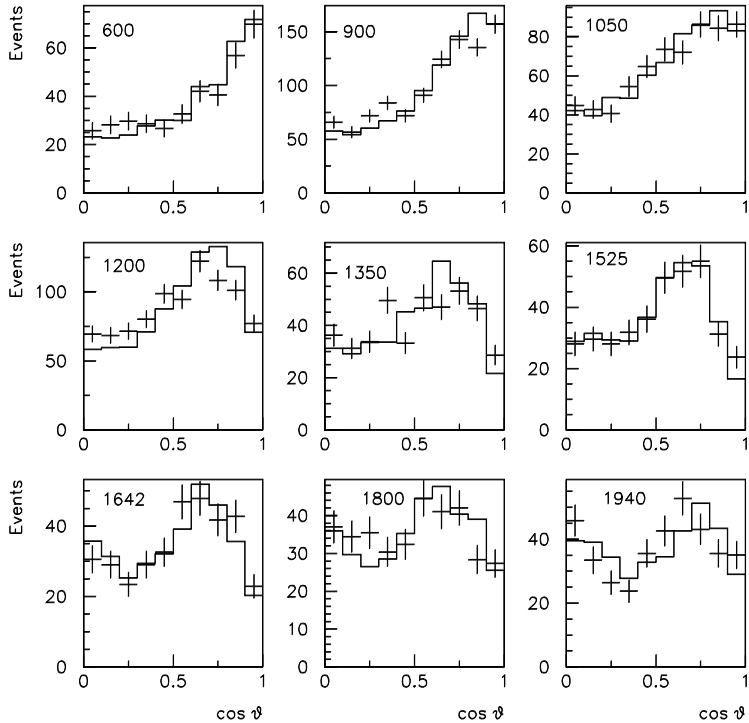,width=11cm}\
\vskip -110.45mm
\hskip 2.9mm
\epsfig{file=FEPE6.EPS,width=11cm}\
~\
\vskip -8mm
\caption{Angular distributions for production of $a_0(980)$ for events
in the $\eta \pi$ mass range 960 to 1020 MeV; histograms show the
partial wave fit. Numbers in each panel indicate beam momenta in MeV/c.}
\end{center}
\end{figure}

We turn now to quantum numbers $J^P = 2^-$.
In an earlier publication on the channel $\eta \eta \pi ^0 \pi ^0$ [8],
we reported evidence for a $\pi _2$ with $M = 1880 \pm 20$ MeV,
$\Gamma = 255 \pm 45$ MeV, decaying dominantly to $a_2(1320)\eta$.
Here, we again find a strong $a_2(1320)\eta$ S-wave contribution
shown by the full curve of Fig. 4(d);
the $2^- \to [a_2\eta ]_{L = 0}$ amplitude is distinctive because of its
$P_2(\cos \alpha )$ dependence on the decay angle $\alpha$ for $a_2(1320) \to
\eta \pi$.
The fitted mass and width for this amplitude are closely consistent
with those for $\pi _2(1880)$. The contribution to $f_0(1500)\pi$,
shown by the dashed curve in Fig. 4(d), is consistent at low masses
with the small branching ratio observed for $\pi _2(1880)$
to that channel in Ref. [8].

However, we observe here an additional
$a_0(980)\eta$ contribution with $J^P = 2^-$,
considerably larger than that allowed in $\eta \eta \pi ^0 \pi ^0$ data of
Ref. [8]. It would be conspicuous there, but is absent. Its fitted magnitude
in present data is shown by the dotted curve in Fig. 4(d).
Table 2 shows that it makes a very large improvement in log likelihood,
namely 531.
Both $a_0(980)\eta$ and  $f_0(1500)\pi$ signals for $J^P = 2^-$
are clearly
recognisable from their production amplitude $P_2(\cos \theta )$,
which interferes distinctively with the large $0^-$ amplitudes;
this interference is clearly visible at the lowest momenta in
the angular distribution against production angle $\theta$
in Fig. 5.

If the $2^- \to [a_0\eta ]_{L = 2}$ channel is fitted freely,
it optimises at $M = 1990 \pm 30$ MeV with a width of $290 \pm 60$ MeV.
The variation of log likelihood with mass is shown in Fig. 4(h).
Errors quoted for mass and width allow for the possibility of moving an
$a_0\eta$ contribution into $\pi _2(1880)$ consistent with the upper limit
from Ref. [8].
If the mass and width are set to those of $\pi _2 (1880)$, despite the
evidence in Ref. [8] against decays of $\pi _2(1880)$ to $a_0(980)\eta$,
log likelihood is worse by 31.
For two degrees of freedom, this is more than a $7\sigma$ effect.

There is further evidence for two distinct $\pi _2$ states from the
$3\pi ^0$ data.
They require a strong $f_2(1270)\pi$ amplitude with $ L = 2$,
consistent in mass and width with $\pi _2(1880)$.
The $f_2(1270)\pi $ amplitude with $L = 0$ peaks at
a higher mass $2020 \pm 17$ MeV with $\Gamma = 165 \pm 35$ MeV.
The width is considerably less than found here.
However, both those data and $\eta \eta \pi ^0$ are well fitted with
a compromise mass $M = 2005 \pm 15$ MeV and width $200 \pm 40$ MeV,
together with $\pi _2(1880)$.

We have earlier found similar evidence for two neighbouring $I = 0$
$J^P = 2^-$ states, $\eta _2(1860)$ and $\eta _2(2030)$ [9].
The $\eta _2(1860)$ has been confirmed by WA102 [10].
Two neighbouring $J^P = 2^-$ states with $I = 1$ and different decay modes
are then a clear possibility.
If the spacing in mass squared follows the empirically observed
1.143 GeV$^2$, the radial excitation of $\pi _2(1670)$ is expected
at $1985 \pm 20$ MeV.
Because two separate $\pi _2$ candidates in a narrow mass range require
one of them to be an intruder state (probably the predicted hybrid), and
because both lie at the bottom of our available mass range, we acknowledge
that confirmation is desirable.
We present the evidence so that other experimental groups should
be alert to the possibility of two separate states.
Present data are consistent with $\pi _2(1880)$ accounting for
the entire $a_2(1320)\eta$ and $f_0(1500)\pi$ signals;
however, we cannot rule out the possibility 
that these two channels are also fed in the present data 
partly by the second state at 2005 MeV.

For other quantum numbers, the $\eta \eta \pi ^0$ data are consistent with
resonances required by the analysis of $3\pi ^0$.
There is a small but highly significant $3^+$
peak at $2260 \pm 50$ MeV in the dashed curve in Fig. 4(e).
Despite large errors for mass and
width, this is additional evidence that a $3^+$ resonance exists in
this mass range.
Likewise, there is a distinct peak in the $2^-$ amplitude for
production of $a_2(1660)\pi$, shown by the chain curve of Fig. 4(d);
unfortunately, the determination of mass and width are poor, both here
and in $3\pi ^0$ data.
There is a distinct $4^-$ signal in $a_0(980)\eta$; Table 2 shows that
it produces a large improvement in log likelihood of 213.
It optimises at $2255 \pm 30$ MeV in present data, very close to the
value $2250 \pm 15$ MeV for $3\pi ^0$ data, where there is a large
$4^-$ signal.
For $J^P = 1^+$, shown by the full curve in Fig. 4(e), there is a definite
signal at the higher masses; this is evidence in favour of the
state required in the analysis of $3\pi ^0$ data at 2270 MeV.
At the bottom of the mass range, a second $1^+$ state gives a small
improvement of 37 in log likelihood.
This improvement is sufficient to require some additional low mass
contribution, but is not sufficient to determine the mass and
width of any possible resonance.
Table 2 includes small contributions from $2^+$ states at 1950, 2030 and
2175 MeV; those states are required by $3\pi ^0$ data, but contributions
to $\eta \eta \pi ^0$ are too small to help determine masses and widths.

A distinctive feature of the $\eta \eta \pi ^0$ data from 1350 to 1940
MeV/c is that they require a strong contribution from a
broad $f_2(1980)$ decaying to $\eta \eta$, channel (7).
It is clearly visible by eye in the angular distributions for decay to
$\eta \eta$, see Fig. 4 of  Ref. [2].
The data required the curious property that the $f_2(1980)$ is produced almost
purely with spin projection $m' = \pm 1$ along the beam direction
in the final state.
The present analysis confirms this result but clarifies the reason
for the helicity dependence.

%Fig. 6
\begin{figure}
\begin{center}
\vskip -24mm
\epsfig{file=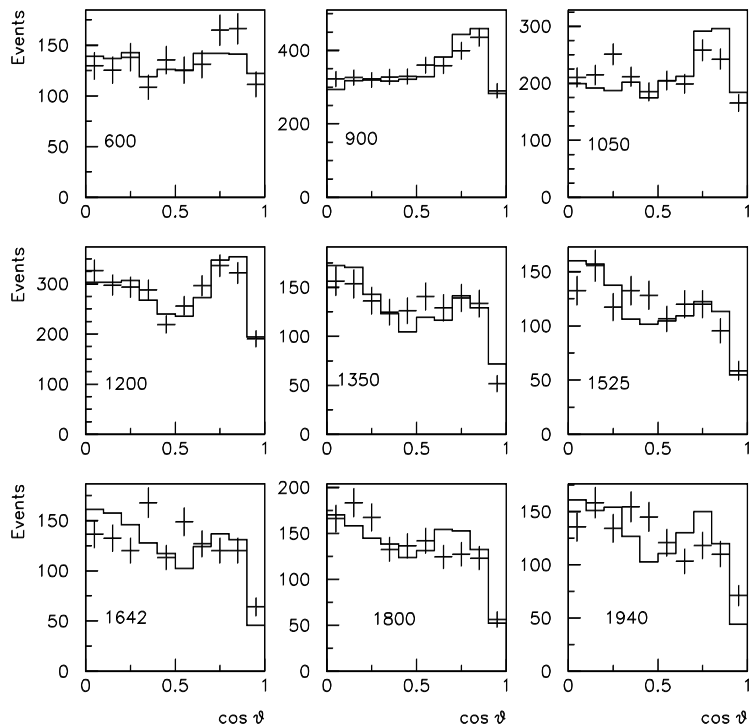,width=11cm}\
\vskip -110.45mm
\hskip 2.9mm
\epsfig{file=FEPE7.EPS,width=11cm}\
~\
\vskip -8mm
\caption{Angular distributions for production of $f_0(1500)$ in the
overall centre of mass for events lying within $\pm 60$ MeV of the
resonance; histograms show the partial wave fit.
Numbers in each panel indicate beam momenta in MeV/c.}
\end{center}
\end{figure}

The large contribution from this state is shown by the dotted curve
on Fig. 4(e).
It peaks at $\sim 2265$ MeV, and is well fitted by an $a_2(2255)$
which appears strongly in the analysis of $3\pi ^0$, $\pi ^0\eta$ and
$\pi ^0\eta '$  data.
We find that the  broad $f_2(1980)$ is produced purely
by this $a_2(2255)$ decaying to $f_2(1980)\pi$ with $L = 1$.
It has a ratio of coupling constants
$g$ between $^3F_2$ and $^3P_2$ amplitudes of
$r_2 = g(^3F_2)/g(^3P_2) = -1.9 \pm 0.4$ for present data;
this ratio agrees well with the value $-2.13 \pm 0.20$
determined in Ref. [1].
With this value of $r_2$, the final state will be
almost purely $m' = \pm 1$ for the following reasons.
Clebsch-Gordan coefficients for coupling of $\bar pp$ to $^3F_2$ and
$^3P_2$ are such that the initial state will be purely $m = 0$ if
$r_2 = -\sqrt {7/2} \simeq -1.9$. The $m = 0$ state decays purely
to final states with $m' = \pm 1$, again because of Clebsch-Gordan
coefficients. Thus, the curious property that the $f_2(1980)$ is produced
almost purely with $m ' = \pm 1$ seems
to be a fortuitous consequence of the fact that the $a_2(2255)$
has $r_2$ close to $-1.9$.

Figs. 5 and 6 show production angular distributions for
$a_0(980)\eta$ and $f_0(1500)\pi$, selecting events within
one half-width of the resonance mass. All are fitted quite well.
Production and decay angular distributions for $a_2(1320)\eta$
are illustrated in our previous publication [3] and are also well fitted.

In summary, a partial wave analysis of $\eta \eta \pi ^0$ data
gives masses and widths for $s$-channel resonances consistent
with those found in the analysis of $3\pi ^0$ data, $\pi ^0\eta$
and $\pi ^0 \eta '$ [1].
For $J^P = 0^-$, the $\eta \eta \pi ^0$ data give the best
determination of the mass and width of the state at 2360 MeV.
An additional $0^-$ state at lower mass is also
required; data from $3\pi ^0$ and $\eta \eta \pi ^0$ are both well
fitted with an average mass of $2070 \pm 35$ MeV.
There is definitely a strong $J^P = 2^-$ amplitude in the $a_0(980)\eta$
channel at low masses, much stronger than observed in Ref. [8]
for $\pi _2(1880)$. It suggests a second $2^-$ state
at $\sim 2005$ MeV, but needs confirmation because both $2^-$
states are near the bottom of the available mass range.

\section{Acknowledgement}
We acknowledge financial support from the
British Particle Physics and Astronomy Research Council (PPARC).
The St. Petersburg group wishes to acknowledge financial support from
PPARC and INTAS grant RFBR 95-0267
We thank Prof. V.V. Anisovich for discussions and helpful criticism.

\end{document}